\documentstyle[prl,aps,epsf,multicol]{revtex} 
\input epsf
\newcommand{\bb}{\begin{equation}}
\newcommand{\en}{\end{equation}}               
\draft 

\begin{document}

\makeatletter
\renewenvironment{table}
  {\let\@capwidth\linewidth\def\@captype{table}}
  {}

\renewenvironment{figure}
  {\let\@capwidth\linewidth\def\@captype{figure}}
  {}
\makeatother

\title{Novel Electrostatic Attraction from Plasmon Fluctuations}

\author{A. W. C. Lau$^{1}$, Dov Levine$^{2}$, and P. Pincus$^{1,3}$ \\}
\address{$^{1}$Department of Physics, 
University of California Santa Barbara, CA 93106--9530\\
$^{2}$ Department of Physics, Technion-Israel Institute of Technology, Haifa 
32000, Israel\\
$^{3}$ Department of Materials, 
University of California Santa Barbara, CA 93106--9530}
\date{\today}
\maketitle

\begin{abstract}
In this Letter, we show that at low temperatures, 
zero-point fluctuations of the 
plasmon modes of two mutually coupled 2-D planar Wigner crystals 
give rise to a novel long-range attractive force.  For the case where the 
distance $d$ between two planar surfaces is large, this attractive 
force has an unusual power-law decay, 
which scales as $d^{-7/2},$ unlike other fluctuation-induced forces. 
Specifically, we  note that its range is longer than the 
``standard'' zero-temperature van der Waals interaction.  
This result may in principle be observed in bilayer electronic 
systems and provides insight into the nature of correlation effects 
for highly charged surfaces.
\end{abstract}

\begin{multicols}{2}

Fluctuation-induced forces are ubiquitous in nature (for a recent review, 
see \cite{fluct}) and constitute an important contribution to the 
interactions of many condensed matter systems\cite{dis}.  
The classic example is the Casimir effect\cite{casimir} in which quantum 
fluctuations of the electromagnetic field between two parallel 
conducting plates lead to an attractive force between them.  
In the context of statistical physics, Fisher and de Gennes
\cite{fisher} have suggested that a similar effect also exists at or 
near the critical point of a system which is confined between two planes.
Other examples include colloid particles immersed in a critical 
fluid\cite{col}, superfluid films\cite{super}, 
liquid crystals\cite{liquid}, and protein inclusions in fluctuating 
membranes\cite{inclusion}.  In general, fluctuation-induced
forces arise because external constraints modify the fluctuations 
of a correlated medium.   These interactions, which are usually 
long-ranged, are controlled by thermal fluctuations at finite 
temperature or quantum fluctuations at low temperature.
In this Letter, we present arguments for a long-range attraction,
derived from the zero-point fluctuations of the {\it plasmon} modes
associated with two 2-D Wigner crystals\cite{wigner}.  

Recent attention has focused on fluctuation-induced forces 
in physical systems that contain charges on surfaces.  
For example, correlation effects in some 2-D electronic systems
in semiconductor heterostructures\cite{quantum} -- 
specifically bilayer systems -- could be viewed as a problem 
of charge-fluctuations on surfaces.  Charge-fluctuation-induced 
forces at high temperatures may have another interesting realization 
in a collection of charged macroions in an aqueous solution of 
neutralizing counterions, with or without added salt\cite{charge,surfaces}.  
The macroions may be charged membranes, stiff polyelectrolytes such as DNA, 
or charged spherical colloidal particles.  
If the surface of macroions is highly charged, 
most of the counterions in solution condense 
onto the surfaces\cite{manning} and their fluctuations
may become important\cite{rods}.  

Specifically, consider a planar surface with charge density $en$,
where $e$ is the electronic charge.  A neutralizing 
counterion in the solution experiences an (unscreened) 
electrostatic attractive force of magnitude
$4\pi n l_B k_B T$, where $l_B \equiv \frac{e^{2}}{\epsilon k_{B}T} 
\approx 7\,$\AA$\,\,$is the Bjerrum length for an aqueous 
solution of dielectric constant 
$\epsilon = 80$ ($H_{2}O$), $k_B$ is the Boltzmann constant, 
and $T$ is the temperature.  The 
length scale -- the Gouy-Chapman length -- at which the thermal 
energy balances the electrostatic energy, given 
by $\lambda = 1 / (4\pi l_B n)$, defines a layer within which
most of the counterions are confined.  For sufficiently 
high charge densities $\lambda \ll L$, where $L$ is the linear size of 
the macroion, the ``condensed'' counterions can be considered as a 
quasi-two-dimensional ideal gas of density $n$.  
To capture correlation effects at high temperature,
it is sufficient to consider in-plane fluctuations about a 
uniform charge distribution.  Explicit calculations show that
charge fluctuations lead to an attractive force between two such 
plates, which scales as $d^{-3}$ for large separations $d$\cite{surfaces}.
This picture may explain the attractive interaction between
two highly charged macroions, observed in experiments\cite{expt} 
and in simulations\cite{numerics}.  Note that
this charge-fluctuation-induced force is 
similar in spirit to the van der Waals interaction, and indeed 
formally identical to the Casimir force between two partially 
transmitting mirrors at high temperatures\cite{mirror}.
Although the available experimental results are for high temperature, 
it is of fundamental interest to understand the 
low temperature interactions as well. Moreover, although it 
is unlikely to be relevant for macroions, such considerations 
may well impact on solid state systems, such as semiconductor bilayers.  
The purpose of this Letter is to study this charge-fluctuation-induced 
force at zero-temperature.

\begin{figure}
\epsfxsize=3.00in
\centerline{\epsfbox{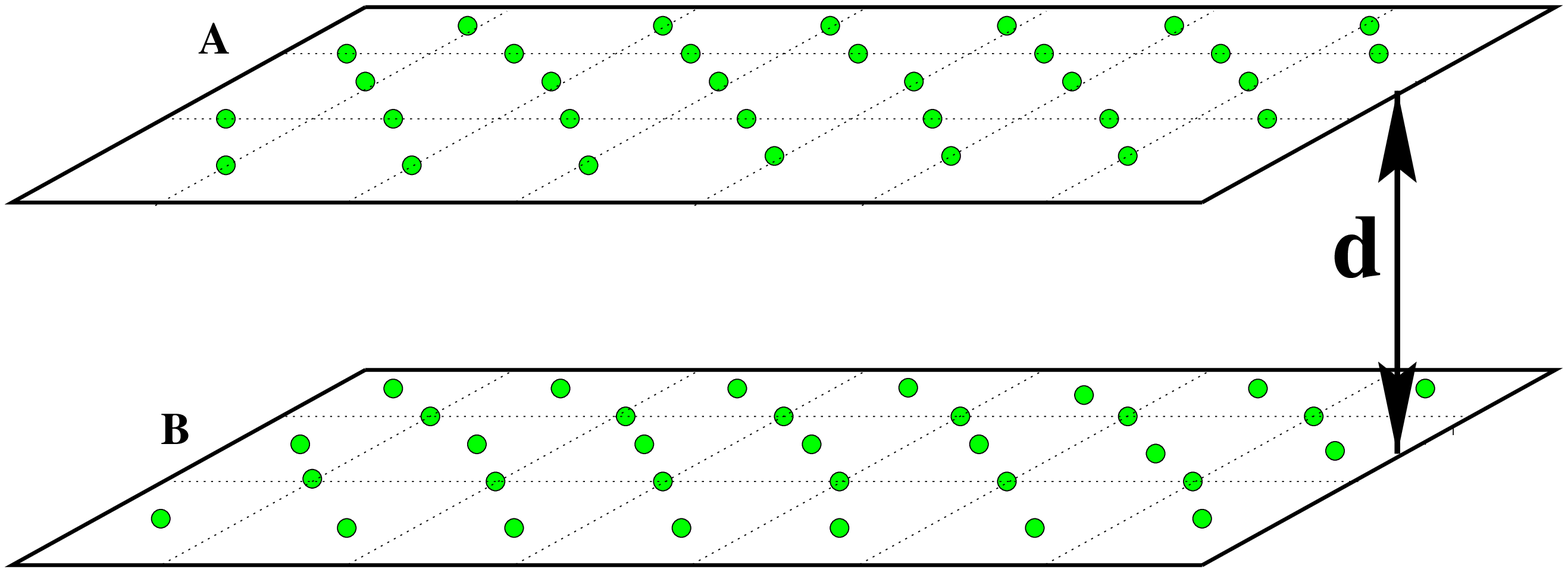}}
\vspace{0.1in}
\caption{Two staggered Wigner
crystals formed by the ``condensed''
counterions.}
\label{wigner}
\end{figure}

To this end, we consider a model system composed of two uniformly 
charged planes, separated by a distance $d$, each having 
a charge density of $e n$. Confined to move on them are negative 
mobile charges, {\em e.g.} classical electrons, with 
particle density $n$ on each of the
surfaces, so that the system as a whole is neutral.
At sufficiently low temperature, 
the charges on the surface crystallize into a 2-D triangular lattice 
as in Fig. \ref{wigner}.  This occurs when 
$\lambda$ is small compared to the average spacing between charges: 
$\lambda \sqrt{n} \ll\,1$, or  $\,\Gamma = 
e^2 \sqrt{\pi n}/\epsilon k_B T$ -- the ratio of 
average Coulomb energy among charges to thermal energy -- is
sufficiently large\cite{background}. 
For electrons on surface of liquid helium, 
it has been experimentally determined that 
$\Gamma \sim 100$\cite{he}.

\vspace{0.1in}

To estimate the attractive force arising from correlation effects
at low $T$, we follow Ref.\cite{lattice} and consider first
the limit of $\Gamma \rightarrow \infty$, {\em i.e.}, the ground state of 
the system.  In this limit, it is easy to see that when two classical 
Wigner crystals experience each other's electric 
fields, the two lattices of charges stagger in order to 
minimize the energy of the system.  In this case,
the ``classical'' electrostatic force per unit area of such a 
charge pattern can be calculated,
\begin{eqnarray}
F_s &(d)&/A  = - \frac{\partial}{\partial d} \left 
\{  \frac{e^2 n}{\epsilon}
\sum_{{\bf R}} \frac{1}{{\sqrt{ |{\bf R} + {\bf c}|^2 
+ d^2}}} \right. \, \nonumber \\
&-& \left. \vphantom{\sum_{{\bf R}} \frac{1}{{\sqrt{ |{\bf R} + {\bf c}|^2 
+ d^2}}}} \frac{(en)^2}{\epsilon } \int 
\frac{d^2{\bf r}}{\sqrt{{\bf r}^2 + d^2}} \right \}
\approx -\frac{2 \pi (en)^2 }{\epsilon}\,e^{-\sqrt{n} d},
\label{static}
\end{eqnarray}
for large $d$; where ${\bf c}$ is the relative displacement 
vector between two lattices of the different plane.
Hence there is a short-range attractive force, which decays exponentially 
with the lattice constant as the characteristic length scale.
In addition to this short-ranged force, 
we show below that zero-point fluctuations 
of the plasmon modes (charge fluctuations) also lead to an attractive 
long-range interaction, analogous to, but scaling differently  (in 
distance) from the standard zero-temperature Casimir effect.  
Moreover, at sufficiently low $T$, this force would still be manifested 
in the system of classical electrons.

To study the zero-temperature charge-fluctuation-induced force, 
we first evaluate the phonon spectrum of the system of two coupled 
Wigner crystals. The positions of the charges are
\begin{eqnarray}
{\bf r}^A ({\bf R}) &=& {\bf R} + {\bf u}^{A}({\bf R}); \nonumber \\
{\bf r}^B ({\bf R}) &=& {\bf R} + {\bf c} + {\bf u}^{B}({\bf R}),
\end{eqnarray}
where $ {\bf u}^{A,B}({\bf R})$ is the small deviation 
from the equilibrium lattice sites ${\bf R}$.  Within the harmonic 
approximation, the potential may be written as
\begin{eqnarray}
\Delta U = {1 \over 2} \sum_{i= A,B}\,\sum_{{\bf R}, {\bf R}'} K_{\alpha \beta}
({\bf R} - {\bf R}')u^{i}_{\alpha}({\bf R})\,u^{i}_{\beta}({\bf R}') 
\nonumber \\
- \sum_{{\bf R}, {\bf R}'} \Delta_{\alpha \beta}({\bf R} - {\bf R}'- {\bf c})
u^{A}_{\alpha}({\bf R})\,u^{B}_{\beta}({\bf R}'),
\end{eqnarray}
where repeated indices are summed.  Here, 
$K_{\alpha \beta}({\bf R} - {\bf R}') = \delta_{{\bf R},{\bf R}'}
\sum_{{\bf R}''}\,[\,\phi_{\alpha \beta}({\bf R} - {\bf R}'') + 
\Delta_{\alpha \beta}({\bf R} - {\bf R}''- {\bf c})] - 
\phi_{\alpha \beta}({\bf R} - {\bf R}'),$
$\phi_{\alpha \beta}({\bf r}) =
\partial_{\alpha}\,\partial_{\beta} \frac{e^2/\epsilon}{|{\bf r}|};
{\bf r} \neq 0$, and $\Delta_{\alpha \beta}({\bf r}) = 
\partial_{\alpha}\,\partial_{\beta} \frac{e^2/\epsilon}
{\sqrt{ {\bf r}^2 + d^2}}$.  
For a general lattice, the square of the 
phonon frequencies are the eigenvalues of the dynamical matrix
\cite{solid}, which in this case may be written as \bb
{\bf D({\bf k}) } = {1 \over m}\,\left [ \begin{array}{cc} {\bf K({\bf k})} 
& - {\bf \Delta({\bf k})} \\  -{\bf \Delta^{\dag}({\bf k})} & 
 {\bf K({\bf k})}  \end{array} \right ],
\label{dyn}
\en
where $m$ is the mass of the charges and 
$ {\bf K({\bf k})} $ and $ {\bf \Delta({\bf k})}$ are $2 \times 2$ 
matrices whose elements are defined by 
$K_{\alpha \beta}({\bf k}) = \sum_{{\bf R}}\,K_{\alpha \beta}({\bf R})\,
e^{-i {\bf k} \cdot {\bf R}}$ and 
similarly for $\Delta_{\alpha \beta}({\bf k})$. 
Using an expansion in reciprocal lattice space, 
they are explicitly given by:
\begin{eqnarray}
K_{\alpha \beta}({\bf k}) &=& \Delta_{\alpha \beta}({\bf 0})
+ \frac{2 \pi e^2 n}{\epsilon} \left \{ \frac{k_{\alpha}\,k_{\beta} }{k}
\vphantom { \sum_{{\bf G } \neq 0} \left 
[ \frac{({\bf G } + {\bf k})_{\alpha}\,
({\bf G } + {\bf k})_{\beta} }{|{\bf G } + {\bf k}| } - 
\frac{G_{\alpha}\,G_{\beta} }{G} \right ]} \right. \nonumber \\
&+& \left.\sum_{{\bf G } \neq 0} \left [ \frac{({\bf G } + {\bf k})_{\alpha}\,
({\bf G } + {\bf k})_{\beta} }{|{\bf G } + {\bf k}| } - 
\frac{G_{\alpha}\,G_{\beta} }{G} \right ] \right \};
\label{K}
\end{eqnarray}
\begin{eqnarray}
\Delta_{\alpha \beta}&({\bf k})&  = - \frac{2 \pi e^2 n}{\epsilon} 
\left \{ \frac{k_{\alpha}\,k_{\beta} }{k}\,e^{-k d} 
\vphantom{\sum_{{\bf G } \neq 0} \frac{({\bf G } + {\bf k})_{\alpha}\,
({\bf G } + {\bf k})_{\beta} }{|{\bf G } + {\bf k}|}\,
e^{-\,|{\bf G } + {\bf k}| d}\,e^{i\,{\bf G}\,\cdot\,{\bf c}}} \right. 
\nonumber \\
&+& \left.\sum_{{\bf G } \neq 0} \frac{({\bf G } + {\bf k})_{\alpha}\,
({\bf G } + {\bf k})_{\beta} }{|{\bf G } + {\bf k}|}\,
e^{-\,|{\bf G } + {\bf k}|\, d}\,e^{i\,{\bf G}\cdot{\bf c}} \right \}, 
\label{delta}  
\end{eqnarray}
where ${\bf G}$ are the reciprocal lattice vectors.  In general, 
to obtain $\omega_j({\bf k})$, the frequency of the 
$j$th mode ($j = 1,...,4$), we have to diagonalize ${\bf D}({\bf k})$
numerically.  This calculation has been performed 
in Ref. \cite{Bi}.  However, since we are interested in the 
long-wavelength limit and large distance asymptotics, we 
approximate ${\bf D}({\bf k})$ in the following fashion:
\begin{eqnarray}
{\bf D({\bf k}) } &\cong&  {1 \over m}\,\left [ \begin{array}{cc} 
{\bf \Delta({\bf 0})} 
& - {\bf \Delta({\bf 0})} \\  -{\bf \Delta^{\dag}({\bf 0})} & 
 {\bf \Delta({\bf 0})}  \end{array} \right ] \nonumber \\
&+& \frac{2 \pi e^2 n}{m \epsilon} 
\,\left [ \begin{array}{cc} 
{\bf D}^{0}({\bf k})
& -{\bf D}^{0}({\bf k})\,e^{-\,k d} \\  -{\bf D}^{0}({\bf k})
\,e^{-\,k d}  & 
 {\bf D}^{0}({\bf k})  \end{array} \right ]
\label{appD}
\end{eqnarray}
where we have defined the matrix ${\bf D}^{0}({\bf k})$ with elements
$D^{0}_{\alpha \beta}({\bf k}) = \frac{k_{\alpha} k_{\beta} }{k}.$
This approximation entails neglecting contributions from 
higher order terms in $k$ and from nonzero reciprocal lattice vectors, 
which are exponentially small ($\sim e^{-\,Gd}$).  
Therefore, Eq. (\ref{appD}) is a good approximation 
provided that $d$ is larger than the average spacing 
between charges on the surface.  

Within this approximation, ${\bf D}({\bf k})$ can be 
diagonalized to yield the following dispersion relations:
\begin{eqnarray}
\omega_1^2(k) &=& 2\Delta;\;\;\omega_2^2(k) = 
2\Delta + \frac{2 \pi e^2 n}{m \epsilon} k\,(\,1 - e^{-k d}\,); \nonumber\\
\omega_3^2(k) &=& 0;\;\;\;\;\;\omega_4^2(k) = 
\frac{2 \pi e^2 n}{m \epsilon} k\,(\,1 + e^{-k d}\,),
\end{eqnarray}
where we have chosen ${\bf \Delta}({\bf 0})$ 
appropriate for staggered triangular lattices: $\Delta_{11} = 
\Delta_{22} = \Delta$ and $\Delta_{12}= \Delta_{21}= 0$, 
with $\Delta \sim e^{-G d}$\cite{delta}.  These modes can also be 
derived by treating the coupling between two isolated Wigner crystals 
as a perturbation.  Mode 1 is one of the two optical modes 
which correspond to out-of-phase vibrations 
of the charges in opposite planes.  The finite gap 
at ${\bf k} = 0$ vanishes exponentially with $d$ for 
large distances as also found in Ref. \cite{Bi}.
Mode 3 is the transverse phonon mode, which 
describes the shear mode of the system, similar to that
of a single 2D Wigner crystal.  We remark 
that this approximation gives zero for its frequency, 
but upon including terms that involve next order in $k$
the dispersion is linear: $\omega_3(k) \sim v_s k$\cite{Bi}.  
Its sound velocity $v_s$ is roughly a constant -- the transverse sound 
velocity for an isolated Wigner crystal -- plus a small correction 
(exponentially decaying for large $d$) which arises from 
the interlayer coupling.  In fact, all higher 
order terms are exponentially damped for 
large $d$.  Modes 2 and 4 may be interpreted as the out-of-phase 
and in-phase plasmon modes, 
respectively.  An interesting feature is that the out-of-phase 
plasmon mode has a gap at $k = 0$ in the presence 
of the coupling.  The in-phase plasmon mode vanishes
as $\sqrt{k}$ as $k \rightarrow 0$ and its sound
velocity diverges.  Physically, the transverse phonon and 
the in-phase plasmon mode describe the charges in different planes 
oscillating in-phase.  We note that their existence has been shown to be 
a general property for a 2D Coulomb plasma\cite{modes}, and
does not specifically depend on the nature of the underlying 
Wigner lattice.  Thus, the results of this paper are universal, 
independent of the nature of the ground state.

For the zero point energy, the dominant modes in the
long wavelength limit are the plasmons: modes 2 and 4.
Neglecting exponentially small contributions, 
{\it i.e.} $\Delta \rightarrow 0$, we obtain the 
zero-point energy (relative to the infinite separation) 
associated with the interactions between 
the surfaces
\begin{eqnarray}
\Delta E_0 &=&\frac{\hbar}{2} \sum_{{\bf k}, j}\,\, 
[\, \omega_{j}({\bf k}, d) - \omega_{j}({\bf k}, d \rightarrow \infty)\,]
= A \,\sqrt{ \frac{n \pi  e^2 \hbar^2}{2 m \epsilon}} \nonumber \\
&\times &\,\int\,\frac{d^2{\bf k}\,\sqrt{k}}{(2\,\pi)^2}\,
\left \{ \sqrt{ 1 + e^{-kd}} + \sqrt{ 1 - e^{-kd}} - 2 \right \},
\end{eqnarray}
where $A$ is the area of the planar surface.  This leads to an attractive 
pressure\bb
\Pi_0(d) = - \frac{1}{A} \frac{\partial\,\Delta E_0(d)}{\partial\,d}=
- \sqrt{ \frac{\hbar^2 e^2 n}{m \epsilon}} \frac{\alpha}{d^{7/2}},
\label{pressure}
\en
where $\alpha$ is a positive numerical constant of order unity, explicitly
given by$$
\alpha =  \frac{1}{4\sqrt{2 \pi}}
\int_0^{\infty}\,dx\,x^{5/2}\,e^{-x}\,\left \{
\frac{1}{\sqrt{ 1 - e^{-x}}} - 
\frac{1}{\sqrt{ 1 + e^{-x}}} \right \}.
$$

Thus, zero-point fluctuations induce a long-range attraction  
which decays with a novel power law $\sim d^{-7/2}$.  This should be
contrasted with the usual Casimir-like force $\sim d^{-4}$,
which arises from, for example, the acoustic phonon zero-point fluctuations. 
We note that this power law stems from the 
2 dimensional nature of charged systems: 2-D plasmons do not have a 
finite gap, as they do in 3D.  Note also that all the 
higher order terms in $k$, as well as those that we
have neglected in our derivation, decay exponentially with $d$; 
therefore Eq. (\ref{pressure}) is the dominant term for
large distances.  For an order of magnitude estimate, 
assuming $m \sim 10^{-25}\,kg$, $n \sim 1/50\,$\AA$^{-2}$, $d \sim 10\,$\AA, 
and $\epsilon \sim 80$, we find $ \Pi_0 \sim 10^{-25}\, J/$\AA$^{3}$.
This is close to the magnitude of the ``classical'' force in 
Eq. (\ref{static}): $F_s/A \sim 10^{-24}\,J/$\AA$^{3}$, and thus 
may be just as important under suitable conditions.

To recapitulate, we have argued that there is
a long-range force at $T=0$, derived from the
zero-point fluctuations, which must be added to the zero-temperature 
``classical'' force.  At finite temperatures,
an explicit calculation\cite{lau} using the Bose-Einstein distribution
shows that at large separations, an additional contribution from the 
plasmon modes to the attractive force is of the form 
$\alpha_{2} k_B T d^{-3}$, which agrees exactly (even the
prefactor $\alpha_{2}$) with the high temperatures result of 
Ref. \cite{surfaces}.  Moreover, the effect of finite $T$ 
on the exponential force is to modify it with a 
``Debye-Waller'' factor, weakening it, and
eventually causing it to vanish\cite{lau}. Thus, for finite $T$, 
we have the following expression for the correlated attractions 
for a system of two coupled planar Wigner crystals:\bb
\Pi(d) = -\alpha_{0} e^{-d/a} - \alpha_{1}\,\hbar\,d^{-7/2} - 
\alpha_{2} k_B T d^{-3},
\en
where $\alpha_{0,1,2}$ are constants and 
$a$ is the range of the short-ranged attraction, of the order of 
the lattice constant.  

Finally, we comment that experimental observations of our result
may prove subtle, as indicated by other examples of 
phonon-fluctuation-induced interactions.  For example, in Ref. \cite{cole}
the effect of zero-point fluctuations of phonons
on the wetting transition of a He thin film was investigated, and
the effect was shown to be small.  Perhaps, a more appropriate 
system in which this behavior may be manifested is bilayer quantum 
well systems.  Indeed, recent experimental techniques allow 
for 2-D confinement of electrons, and a 2D plasmon dispersion has been 
confirmed experimentally in GaAs/AlGaAs heterostructures\cite{isihara}, 
and may exhibit fractional-power-law predicted in this Letter.  
Although there has been experimental progress in observing 
single Wigner crystals\cite{well}, the study of interacting 
Wigner crystals remains an experimental challenge.  In any event, 
the attractive force discussed in 
this Letter may assist with the conceptual understanding of 
correlation effects for highly charged surfaces and more
fundamentally, indicates that for two neutral planes 
with mobile charges, the zero-point interaction has a 
universal scaling different from that of the standard Casimir effect. 

We would like to thank Professors Walter Kohn, S. J. Allen, 
and H. A. Fertig for stimulating and helpful discussions.  
AL and PP acknowledge support from NSF grants MRL--DMR--9632716, 
DMR--9624091, DMR--9708646, and UC-Biotechnology Research and
Education Program.  DL acknowledges support from 
Israel Science Foundation grant 211/97.  DL also would 
like to thank the MRL at Santa Barbara for its wonderful hospitality.

\end{multicols}

\begin{thebibliography}{99}
\bibitem{delta}
For other lattices, there are slight differences, but the 
long wavelength behavior remains unchanged.
\end{thebibliography}
\end{document}